\documentclass[amsmath,amssymb,floatfix,prb,epsf,showpacs,twocolumn]{revtex4}
\usepackage{amsmath,amssymb,natbib,bm,graphicx,url,epsfig}
\usepackage[ansinew]{inputenc}
\usepackage{bm}

\newcommand{\be}{\begin{equation}}
\newcommand{\ee}{\end{equation}}
\newcommand{\bes}{\begin{equation}\begin{split}}
\newcommand{\ees}{\end{split}\end{equation}}
\newcommand{\ba}{\begin{eqnarray}}
\newcommand{\ea}{\end{eqnarray}}
\newcommand{\nn}{\nonumber}

 \DeclareMathOperator{\tr}{tr}

\def\beq{\begin{equation}}
\def\eeq{\end{equation}}
\def\bea{\begin{eqnarray}}
\def\eea{\end{eqnarray}}

\begin{document}

\title{Majorana path integral for nonequilibrium dynamics of two-level
    systems
}

\author{Tigran A.~Sedrakyan and Victor M. Galitski}

\affiliation{Condensed Matter Theory Center and Joint Quantum Institute, Department of Physics, University of Maryland, College Park, MD 20742}

\date{\today}

\begin{abstract}

We present a new field-theoretic approach to anaylize non-equilirbium dynamics of two-level systems (TLS), which is based
on a correspondence between a driven TLS and a Majorana fermion field theory coupled to bosonic fields.
This  approach allows us to calculate analytically  properties of non-linear TLS dynamics with an arbitrary accuracy. We apply our method to
analyze specific TLS dynamics under a monochromatic periodic drive that is relevant to the problem of decoherence in Josephson junction qubits. It is demonstrated that the method gives the precise positions of the resonance peaks in the non-linear dielectric response
function that are in agreement with numerical simulations.
\end{abstract}

\maketitle

\section{Introduction}

A driven two-level-system (TLS) represents a canonical dynamical system
that features a rich variety of interesting non-linear phenomena including various resonance effects,
quantum interference phenomena due to level crossings, coherent destruction of tunnelling, etc (see e.g., Ref.~[\onlinecite{Weiss-2008}]).
Despite the formal simplicity of its formulation, this quantum dynamical system does not admit an exact
solution in a closed analytical form for an arbitrary external drive and one often resorts to numerical
simulations or various approximation schemes for its analysis. There are deep mathematical reasons
for the lack of our ability to solve the corresponding differential equations, which go back to the old works
on Riccatti differential equations and Lie theory. The mathematical problem itself has a wide
spectrum of applications in technology and  physics ranging from technologically important nuclear magnetic
resonance spectroscopy~\cite{Van} to Maxwell-Bloch theory of two-level lasers~\cite{Numai}, non-equilibrium superconductivity~\cite{Yuz,Galitski2010},
and fundamental field-theoretical models such as the Wess-Zumino-Witten theory~\cite{WZW}. Furthermore, with the increased technological ability
to fabricate and control quantum systems, new realizations of the model arise on a continuous basis, such as, for example,
 artificial ``atoms'' interacting with strongly oscillating fields~\cite{art}. This variety of new applications has motivated
focused theoretical researches of the model recently, see e.g., Refs.~[\onlinecite{Kleff2004,Nesi2007,Hausinger2010,ADG}]
 that studied a weak driving limit with small number of photons and Ref.~[\onlinecite{Blais2004}] that investigated the regime of
 strong driving with large photon numbers, just to name a few relevant papers.

Another important field, where the problem of driven TLS dynamics is of great importance, is quantum computing.
There are various realizations of qubits~\cite{Mooij2005,Nakamura1999,Mooij1999},
which in the course of quantum evolution may exhibit interesting non-linear phenomena, such as interference between multiple Landau-Zener transitions  at a level crossing, where adiabatic evolution between them results in an oscillatory qubit magnetization in the
regime of strong qubit driving~\cite{rudner,harmonic}. Physics of driven TLS shows up in qubits
also from a different perspective: Low-energy charge defects are widely believed to be the dominant source
of dephasing in superconducting Josephson junction qubits~\cite{MH}. There, interactions between the charged TLS defect and an applied electric field gives rise to the same problem of a non-equilibrium TLS under a periodic monochromatic drive.
In general driven TLS was a subject of intensive studies during the last decade, different aspects of which
are presented in e.g., Refs.~\onlinecite{Frasca, Ashhab, Son, Grifoni-2, Baraats}.

In most cases mentioned above, the basic problem that we are actually interested in is summarized by the simple
Hamiltonian, $H=-{\bf \Delta}(t) \cdot \hat{\bm \sigma}$, which leads to the following evolution operator $\hat{U}(\tau)$ and
the ``partition function''  ${\mathcal Z}$

\bea
\label{rep}
\hat{U}(\tau)= \large{\hat{T}}
\exp\left\{i \int_0^{\tau}d t {\bf \Delta}(t)\cdot \hat{\bm \sigma}\right\}, \quad  {\mathcal Z}=\tr \hat{U}(\tau).
\eea
Here $ \hat{T} $ is the time ordering operator, $\hat{\bm \sigma}=(\hat{\sigma}_1,\hat{\sigma}_2,\hat{\sigma}_3)$ is the  vector of
Pauli matrices, and ${\bf \Delta}(t)=\left[\Delta_1(t),\Delta_2(t),\Delta_3\right(t)]$ is the three-component external driving field.

In this paper, we propose an analytical Majorana field-theory approach,
which is capable of describing {\em arbitrary} external driving fields.
Though the approach can be applied to the most general case however we consider following
form of the periodic driving field:
${\bf \Delta}(t) = \left[ \Delta_1(t), \Delta_2(t), \Delta_3 \right]$, where $\Delta_3$ is time independent constant
(this assumption does not break the generality as it corresponds to transforming to a frame with static $\Delta_3$) and we also assume that the two time-varying components
 have a finite number of harmonics, $\Delta_i(t)=\epsilon_i+\sum_nA_n\cos(\omega_n t)$, $i=1,2$. Let us first summarize three steps that
 one has to follow to analyze this quantum dynamical system within our approach: (1)~Fourier transform the fields:
\bea
\label{FC}
\Delta_i(\omega)=\epsilon_i\delta(\omega)+\sum_{n,\sigma=\pm}A_n\delta(\omega+\sigma\omega_n);
\eea
(2)~Write exact Dyson equations for the fermionic Green's function, ${\mathcal K}(\omega,\omega^\prime)$ in a Majorana field theory; (3)~Solve the resulting equations recursively.
The following presentation deciphers this prescription and provides a specific example of its use for the most experimentally relevant (but analytically unsolvable) case of a simple monochromatic drive:
\begin{eqnarray}
\label{1}
\Delta_1(t)= \frac{\varepsilon}{2}+A \cos (w t),\; \Delta_2(t)\equiv 0,\;  \Delta_3(t)=\frac{\Delta}{2}=\text{const},
\nn\\
\end{eqnarray}
were $\varepsilon$ is an energy splitting, $w$ and $A$ are the frequency
and amplitude of an external field.
Our main result is a practically-useful continued fraction representation  of the
correlation function, $K(t,t^\prime)=\tr \langle\hat{\sigma}_3(t)\hat{\sigma}_3(t^\prime)\rangle$,
that contains its exact spectrum.

\section{Action in terms of Majorana fermions}

Our work is based on the observation that the expression
 for the ``partition function,'' ${\mathcal Z}$, resembles the Green's function of a
 spinning particle passing in a one-dimensional space from point $x(t=0)$ to $x(t=\tau)$
 in Feynman path integral quantization approach~[\onlinecite{Polyakov}], where ${\bm \Delta}(t)$ has a meaning of
 the velocity, $\dot{\bf x}(t)$, of the spinning particle.
Subsequently, the conversion to Majorana fields is achieved by using the
 approach of Refs.~[\onlinecite{Polyakov,BM}], which  provides a simple prescription:
 the Pauli matrices
$\hat{\sigma}_{\mu},\; \hat{\sigma}_3=i \hat{\sigma}_1\hat{\sigma}_2 $ should be replaced by the Majorana fields
$\xi_{\mu}(t),\; \xi_3(t)$ respectively.
 In other words, Pauli matrices
can be regarded as a quantized version of path-integral Majorana fields.
We note that Majorana fermion is
 its own antiparticle,
{\it i.e.}, its creation and annihilation operators are identical.
As fields, they can be described by real valued Grassmann variables~\cite{Berezin,BM} denoted below as $\xi(t)$.
The spin dynamics has been investigated
in a different context using Majorana fermion representation
in Ref.~[\onlinecite{col1}]. Similar approaches have been  employed
previously in Refs.~[\onlinecite{VP,Tsv,Sachdev,TSV1,Col3,V81,Shastry}].



A cornerstone of this work is that the ``partition function,'' ${\mathcal Z}$, i.e., the
trace of the evolution operator can be exactly reproduced within a field theory of three Majorana fermions $\xi_1(t)$, $\xi_2(t)$, and $\xi_3(t)$, defined by the
functional integral
\bea
\label{ZZ}
{\mathcal Z}= \frac{1}{{\cal Z}_0}\int {\cal D}\xi_1(t){\cal D}\xi_2(t){\cal D}\xi_3(t) e^{i {\mathcal S}\bigl(\{\xi\},{\bf \Delta}\bigr)}.
\eea
where ${\cal Z}_0=Det\left[\frac{d}{dt}\right]$, and  the action [here we set $\Delta_3(t) \equiv \Delta/2={\rm const}$] has the form:
\bea
\label{act1}
\!\!\!\! i {\mathcal S}\bigl(\{\xi\},{\bf \Delta}\bigr)&=&\int_0^{\tau}dt\;\left[\frac{1}{4}\xi_\mu(t)\dot{\xi}_\mu(t)
+\frac{1}{4}\xi_3(t)\dot{\xi}_3(t)\right. \nn\\
&+&\Delta_\mu(t)\xi_\mu(t)\xi_3(t)
-\left. \frac{1}{4}\epsilon_{\mu\nu}\xi_\mu(t)\xi_{\nu}(t)\Delta\right].
\eea
In Eq.~(\ref{act1})  $\mu,\nu=1,2$, $\epsilon_{12}=-\epsilon_{21}=1$, $\epsilon_{11}=\epsilon_{22}=0$ and the summation over repeated indices is
implied.
Equivalence of the driven TLS (\ref{1}) and the Majorana field theory (\ref{act1}) can be shown in three steps:

(i) First we expand the exponent of the interaction term in the expression of the partition
function, ${\mathcal Z}$ [see Eq.~(\ref{ZZ})]
\bea
\label{i1}
{\mathcal Z}&=\frac{1}{{\cal Z}_0}&\sum_{n=0}^{\infty}  \int {\cal D}{\bm \xi}(t) {\cal D}\xi_3(t) \int_0^{\tau}\!dt_1 \int_0^{t_1}\!dt_2 \cdots
\int_{0}^{t_{n-1}}\!dt_n \nn\\
&&\big(\Delta_\mu(t_1)\xi_\mu(t_1)\xi_3(t_1)
- \frac{1}{2}\epsilon_{\mu\nu}\xi_\mu(t_1)\xi_{\nu}(t_1)\Delta_3\big)\cdots \nn\\
&&\big(\Delta_\mu(t_n)\xi_\mu(t_n)\xi_3(t_n)
- \frac{1}{2}\epsilon_{\mu\nu}\xi_\mu(t_n)\xi_{\nu}(t_n)\Delta_3\big)\nn\\
&\times&\exp\left\{{\int_0^{\tau}dt\;\left[\frac{1}{4}\xi_\mu(t)\dot{\xi}_\mu(t)
+\frac{1}{4}\xi_3(t)\dot{\xi}_3(t)\right]}\right\}.
\eea

(ii) At the second stage we make  use of the free Green's
function of the Majorana fields (see details in Appendix A)
\bea
\label{identity}
\langle\xi_{\mu}(t_1) \xi_{\nu}(t_2)\rangle &=&\frac{1}{{\cal Z}_0}\int {\cal D}{\bm \xi}(t) \xi_{\mu}(t_1) \xi_{\nu}(t_2)
\\ &\times&\exp \left[\frac{1}{4}\int dt \; \xi_{\rho} \dot{\xi}_{\rho}\right]
=\delta_{\mu\nu}\text{sign}[t_1-t_2],\nn
\eea
This form of the Green's function follows straightforwardly from the identity $\frac{d}{dt}sign[t-t']=\delta[t-t']$.
Then, by use of the Wick's theorem, one can replace correlation functions of the Majorana fields in the expression (\ref{i1}) by the
trace of the product of Pauli matrices. This procedure yields
\bea
\label{i2}
&&\frac{1}{{\cal Z}_0}\int {\cal D}{\bm \xi}(t) \xi_{\mu_1}(t_1) \xi_{\mu_2}(t_2)\cdots \xi_{\mu_n}\nn\\
&\times& \exp\left\{{\int_0^{\tau}dt\;\left[\frac{1}{4}\xi_\mu(t)\dot{\xi}_\mu(t)
+\frac{1}{4}\xi_3(t)\dot{\xi}_3(t)\right]}\right\}
\nn \\
&=&\sum_{all pairings}(-1)^{P}\delta_{\mu_1 \mu_{i_1}}\delta_{\mu_2 \mu_{i_2}}\cdots \delta_{\mu_{n/2} \mu_{i_{n/2}}}\nn\\
&=& Tr\big[\sigma_{\mu_1} \sigma_{\mu_2} \cdots \sigma_{\mu_n} \big],
\eea
where $P$ is the number of permutations needed for obtaining the set of indices
$\mu_1 \mu_{i_1} \mu_2 \mu_{i_2} \cdots \mu_{n/2} \mu_{i_{n/2}}$ from
$\mu_1 \mu_2 \cdots \mu_n$ (note that $n$ here is even). It is also important to note
that $sign[t-t']$ factors in the
Majorana fermion Green's function (\ref{identity}) disappear in the expression  above due to the time ordering
of fields in Eq.~(\ref{i1}).
Similar situation is with the functional integral over the fields $\xi_3$, which satisfy the condition
$\langle \xi_3(t)\xi_3(t') \rangle =sign[t-t']$. This suggests that the fields $\xi_3(t)$ simply can be effectively
dropped from Eq.~(\ref{i2}), as the whole integral gives one.
Finally, the expression (\ref{i2}) shows that one can replace Majorana fields, ${\bm \xi}_{\mu}(t)$,
by the corresponding Pauli matrices, $\sigma_{\mu}$.

 (iii) Upon replacing the fields $\xi_{\mu}(t)$ in Eq.~(\ref{i2}) by the
  Pauli matrices, $\sigma_{\mu}$,
 dropping $\xi_3(t)$ and subsequently collecting the obtained series back
 to the exponent one will obtain ${\cal Z}=Tr \large{\hat{T}}
\exp\left\{i \int_0^{\tau}d t {\bf \Delta}(t)\cdot \hat{\bm \sigma}\right\} $. In derivation of this formula
we also use the relation $\sigma_3=i \sigma_1 \sigma_2$.

As we mentioned above, action (\ref{act1}) can be
interpreted as the  path-integral quantization of relativistic
spinning particle in a fixed gauge
in an external magnetic field $\Delta_3(t)=\Delta/2$ ~\cite{Polyakov}.

Finally, Gaussian integration over the fields $\xi_\mu(t)$, $\mu=1,2$ gives the partition function in the form
${\mathcal Z}=\int D[\xi_3(t)]\exp\left\{iS_3[\xi_3(t),{\bf \Delta}(t)]\right\}$,
where
the effective action for the $\xi_3(t)$ field reads
\bea
\label{S3}
\!\!\!\!\!\!&& i {\mathcal S}_{3}\left[\xi_3,{\bf \Delta}(t)\right] = \frac{1}{4}\int_0^\tau dt \xi_3\dot{\xi}_3
+ 2 \int_0^\tau dt dt^\prime\xi_3(t^\prime)\times\\
\!\!\!\!\!\!&&\Big[ \Delta_+(t^\prime)G_-(t-t^\prime)
\Delta_-(t)
+\Delta_-(t^\prime)G_+(t-t^\prime) \Delta_+(t)\Big] \xi_3(t).\nn
\eea
Here $G_{\pm}(t-t^\prime) = \frac{1}{2}e^{\mp i \Delta\cdot (t-t')} \text{sign}[t-t']$ is the Green's function
of the differential operator $\left( \partial_t \pm i \Delta \right)$, while
$\Delta_{\pm}(t)=\Delta_1(t)\pm i \Delta_2(t)$. Eq.~(\ref{S3})
is one of the new results presented here.

For our purposes it is convenient to express ${\mathcal S}_{3}$ through the
Fourier images of the fields and functions in Eq.~(\ref{S3}).
Then,
in the limit $\tau\rightarrow\infty$, we obtain for ${\mathcal S}_{3}$
\bea
\label{fourier}
{\mathcal S}_{3}\left[\xi_3,{\bf \Delta}(t)\right]=-\frac{\pi}{2}\int d\omega d\omega^{\prime}\xi_3(\omega)
{\mathcal K}^{-1}(\omega,\omega^\prime)\xi_3(\omega^\prime),\nn\\
\eea
where
${\mathcal K}(\omega,\omega^\prime)=\left[\omega\delta(\omega+\omega^\prime)+G(\omega,\omega^\prime)\right]^{-1}$,
and $G(\omega,\omega^\prime)$ is an antisymmetric kernel
 given by
\bea
\label{G}
G(\omega,\omega^\prime)&=&2\int d\omega_1\frac{\Delta_+(\omega_1)
\Delta_-(\omega+\omega_1+\omega^{\prime})(\omega^\prime-\omega)}
{(\omega_1+\omega+\Delta)(\omega_1+\omega^\prime+\Delta)}.\nn\\
\eea

Due to the novelty of our approach, we first briefly derive several established
results and then turn to the main problem of dephasing in the presence of a monochromatic drive.


\section{Application to superconductivity and verification of the BCS result}

The aim of the present section is to apply the developed technique to
study the nonperturbative properties of the pairing Hamiltonian.
The textbook expression of the partition function of the pairing model
is given in terms of a functional integral with respect to complex Grasmann fields,
$c_{{\bf k}\sigma }(t)$, $\bar{c}_{{\bf k}\sigma}(t)$, $\sigma=\uparrow,\downarrow$, with the action including four-fermionic pairing interaction. The standard
approach to treat this action is to decouple the interaction term by introducing
a set of bosonic fields, $\check\Delta(t)$, $\overline{\check\Delta}(t)$, over which
one will have an additional functional integral.
Partition function
corresponding to the zero-particle and paired sectors of the pairing Hamiltonian reduces to~[\onlinecite{Gorkov}]
\bea
\label{BCS}
Z_{BCS}&=&\int {\mathcal D}[\check\Delta, \overline{\check\Delta}, c, c^+] e^{-S_{BCS}},\nn\\
S_{BCS}&=&\int_0^{\beta} d\tau \Big\{\sum_{\bf k}\bar\psi_{\bf k}(\partial_{\tau}+h_{\bf k})\psi_{\bf k}
+\frac{1}{g}\overline{\check\Delta} \check\Delta\Big\}
\eea
where $g$ is the interaction constant,
\bea
\label{nambu}
\psi_{\bf k}= \Biggl(
\begin{array}{c}
c_{{\bf k},\uparrow}\\
\bar{c}_{{\bf -k},\downarrow}
\end{array}
\Biggr)
\eea
defines the Nambu spinor and
\bea
\label{h}
h_{\bf k}=\epsilon_{\bf k}\sigma_3 + \check\Delta_1 \sigma_1 + \check\Delta_2 \sigma_2
\eea
is the matrix Hamiltonian, where $\check\Delta =\check\Delta_1-i \check\Delta_2 , \overline{\check\Delta} =\check\Delta_1+i\check\Delta_2$.
Remarkably, the form of the operator $h_{\bf k}$ is very much reminiscent to our driven Hamiltonian (\ref{1}), but with the third
component, $\epsilon_{\bf k}$, being a time independent constant.
Therefore it is straightforward to employ the above developed technique of Majorana field theory to
represent the exact BCS partition function, $Z_{BCS}$ as a functional with respect to one specie Majorana fermion, $\xi_3(t)$,
and Hubbard-Stratonovich boson fields, $\check\Delta$ and $\overline{\check\Delta}$, see Eq.~(\ref{fourier}).
If one is interested for example in calculation of $Z_{BCS}$, then,
by integrating out $\xi_3(t)$ one produces an effective bosonic action:

%
\bea
\label{pair}
Z_{BCS}=\int{\mathcal D}\check\Delta_\mu e^{S_{eff}(\check\Delta)}, \hspace{4.5cm}\!\!\!\\
S_{eff}=-\int d\varepsilon\frac{{\bm \check\Delta}^2(\varepsilon)}{g}+\frac{1}{2}
\tr \log \left[\omega\delta(\omega+\omega^\prime)+G(\omega,\omega^\prime)\right]\nn
\eea
where $G(\omega,\omega^\prime)$ is a functional of $\check\Delta(t)$ and $\overline{\check\Delta}(t)$,
and is defined by Eq.~(\ref{G}). Note that the functional integral Eq.~(\ref{pair}) is formally
nothing but a nonlinear functional determinant written in energy (rather than imaginary time) space.
Variation of the effective action with respect to the bosonic fields gives
the equation of motion, $\frac{\delta S_{eff}}{\delta\check\Delta_\pm(\tilde\omega)}=0$, which in turn
yields the general gap equation 
\bea
\label{ggap}
-\frac{2\check\Delta_\pm(\tilde\omega)}{g}&+&\frac{1}{2}\int d\omega d\omega^\prime {\cal K}(\omega,\omega^\prime)\\
&\times&\frac{\check\Delta_\pm(\tilde\omega+\omega+\omega^\prime)
(\omega^\prime-\omega)}{(\tilde\omega+\omega+\epsilon_{\bf k})(\tilde\omega+\omega^\prime+\epsilon_{\bf k})}=0,\nn
\eea
written in the energy representation.

The pairing hamiltonian itself is designed to describe the superconducting phase,
where the order parameter, $\check\Delta$, is different from zero.
The gap equation (\ref{ggap}), {\em i.e.} the equation for the Fourier-transformed
bosonic Hubbard-Stratonovich field, $\check\Delta(\omega)$, describes the dynamics of the
order parameter in the global gauge symmetry broken phase.

The BCS mean-field solution corresponds to the choice $\check\Delta_{\pm}(\omega)=\Delta_0^{\pm}\delta(\omega)$,
where  $\Delta_0^{\pm}$ is $\omega$ independent. In other words it assumes that the time-dependence of
the order parameter is unimportant.
Then, from Eq.~(\ref{G}) we find that
\bea
\label{BCS1}
G_{MF}(\omega,\omega^\prime)=\frac{\omega {\bf \Delta}_0^2}{\epsilon_{\bf k}^2+{\bf \Delta}_0^2-\omega^2}\delta(\omega+\omega^\prime).
\eea
and 
\bea
\label{DBCS}
{\cal K}_{MF}(\omega,\omega^\prime)=
\frac{\omega(\epsilon_{\bf k}^2+{\bf \Delta}_0^2-\omega^2)}{\epsilon_{\bf k}^2-\omega^2} \delta(\omega+\omega^\prime).
\eea
Making use of Eqs.~(\ref{BCS1}) and (\ref{DBCS}), and substituting them into (\ref{ggap}), we obtain
\bea
\label{BCS2}
-\frac{2}{g}+N(0)\int d\omega\frac{1}{\epsilon_{\bf k}^2+{\bf \Delta}_0^2-\omega^2}=0,
\eea
which reproduces the standard BCS gap equation with $N(0)$ being the
approximated to a constant density of states. 

\section{Non-dissipative two-level systems}

Neglecting environmental effects on the driven TLS we consider
the Hamiltonian $H= -{\bf \Delta(t)} \hat{\bm \sigma} $, with
the driving fields given by Eq.~(\ref{1}).
Our goal is the calculation of spin-spin correlation function
\bea
\label{spin-cor}
K(t,t')&=&\tr \left[U(-\infty,t)\hat{\sigma}_3 U(t,t^\prime)\hat{\sigma}_3 U(t^\prime, \infty)\right]\nn\\
&=&\tr \left[U^+(t,t^{\prime})\hat{\sigma}_3 U(t,t^\prime)\hat{\sigma}_3\right],
\eea
where $U(t,t')$ is an evolution operator defined in (\ref{rep}).
According to the representation developed above, this function is identical to
the correlation function,  $K(t,t')=\langle \xi_3(t) \xi_3(t^\prime)\rangle$, of the Majorana field, $\xi_3(t)$, calculated with the use of
action ${\mathcal S}_3[\xi_3,{\bf  \Delta}]$ from
Eq.~(\ref{S3}). Importantly, in this formulation, the field $\xi_3(t)$ is the representative of the Pauli matrix $\hat{\sigma}_3$.

An observable of practical interest is the survival probability,
$P_{\downarrow,\downarrow}(t,t')=\left| \langle \downarrow | U(t,t^\prime) |\downarrow \rangle \right|^2$,
which gives the probability of the spin to remain the same as it was in the beginning of evolution at time $t$.
The survival probability is related to the spin-spin correlation function via a
simple identity:
\bea
\label{survival}
P_{\downarrow,\downarrow}(t,t')=\frac{1}{2}\left[ 1+K(t,t^\prime) \right] .
\eea

Before proceeding, it is worthwhile to
calculate the survival probability for $\Delta=0$ and $\Delta_2(t)=0$. In this limit
the correlation function $K(t,t')$ has a rather simple form and can be calculated exactly. First we observe that
one can perform Hubbard-Stratonovich transformation in Eq.~(\ref{S3})  by introducing an additional
Majorana field, $\eta$, and then decouple the quadratic in $\Delta_1(t)$ term. In this new action,
\bea
\label{S3-2}
i \bar{S}_{3}\left[\xi_3(t),\eta(t)\right] =\int_0^\tau dt \Big[ \frac{1}{4}\xi_3\dot{\xi}_3+
 \frac{1}{4}\eta \dot{\eta}+ i \Delta(t)\xi_3\; \eta \Big],\nn\\
\eea
the field $\Delta_1(t)$  plays the role of a "gauge field".
Then we introduce
variables $\xi^{\pm}=(\xi_3\pm \eta)/2$ and eliminate the ``gauge field,'' $i \Delta_1(t)= W^{-1} \partial_t W$
with
$W(t)=\exp[i\int^t dt' \Delta_1(t')]=\exp[i x(t)]$
from the action by rescaling  the  fields,
$\tilde{\xi}^{+}=W \xi^+ $. In this new fields action Eq.~(\ref{S3}) acquires the form of a free
field theory,
$i \bar{S}_{3}\left[\tilde \xi^{\pm}(t)\right] =\frac{1}{2}\int_0^\tau dt \left[\tilde\xi^+\dot{\tilde{\xi}}^-+
\tilde\xi^-\dot{\tilde{\xi}}^+ \right],
$ but the correlation function now transforms into
\bea
\label{d=0}
K(t,t^\prime)&=& 2 \Re\Big[ \langle \tilde{\xi}^+(t)\tilde{\xi}^-(t^\prime)\rangle e^{2 i\int_{t}^{t'} dt' \Delta_1(t')} \Big]\nn\\
&=& \text{sign}[t-t^\prime]\cos\Big[{2 i\int_{t}^{t'} dt' \Delta_1(t')}\Big].
\eea
This known expression can also be obtained directly from the original $T$-ordered exponent.

\begin{figure}[t]
\centerline{\includegraphics[width=70mm,angle=0,clip]{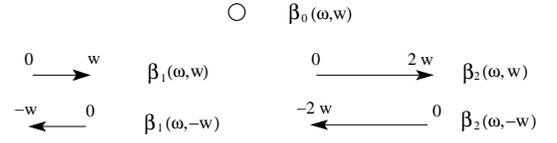}}
\caption{Diagrammatic representation for the matrix elements of $G$. Short [$\beta_1(\omega,\pm w)$] and
long [$\beta_2(\omega,\pm w)$] arrows represent
$\omega \rightarrow \omega \pm w$ and $\omega \rightarrow \omega \pm 2 w$ matrix elements respectively.}
\label{fig:1}
\end{figure}

If $\Delta \neq 0$, it is convenient to perform calculations in
frequency space. From Eq.~(\ref{fourier}),
it follows that the Fourier image of  $K(t,t^\prime)$ is nothing but the
inverse of the operator, $ {\mathcal K}(\omega,\omega^\prime)$.
Therefore, the problem reduces to the calculation of this inverse matrix.

Let us calculate ${\mathcal K}(\omega,\omega^\prime)$ for the
particular  driving field in Eq.~(\ref{1}).
Our strategy will be to firstly calculate $G(\omega,\omega^\prime)$
from its definition (\ref{G}),
by inserting there the Fourier transformed driving field
$\Delta_+(\omega)
=\Delta_-(\omega)= \left[{\epsilon}/2\right] \delta(\omega)+ A \left[\delta(\omega-w)+ \delta(\omega+w)\right]$.
Then, integration over $\omega_1$ yields
\bea
\label{tildeG2}
G(\omega,-\omega^\prime)&=& \beta_0(\omega,w)\delta(\omega-\omega')\\
&+&\sum_{k=1,2; \sigma=\pm} \beta_{k}(\omega,\sigma w)\delta(\omega+ \sigma k w-\omega')\nn
%
\eea
where
\bea
\label{matelements}
 \beta_0(\omega,w)&=&\varepsilon^2 f[\omega]
+A^2 (f[\omega-w]+f[\omega+w]),\;\;\nn\\
\beta_{1}(\omega,w)&=& \varepsilon A (f[\omega]+f[\omega+w]),\;\;\nn\\
\beta_{2}(\omega,w)&=& A^2 (f[\omega-w]+f[\omega+w])
\eea
with
$f[x]=2 x/(x^2-\Delta^2)$.
Secondly, in order to find the inverse of ${\mathcal K}^{-1}$ ({\em i.e.} ${\mathcal K}$), we use the identity
$ {\mathcal K}= \sum_{k=0}^{\infty}(1- {\mathcal K}^{-1})^k$, where $1$ stands for the identity matrix.
This sum formally exist in the region $|1-{\mathcal K}^{-1}|\leq 1$. Outside of this
region the inverse should be understood as the analytic continuation of the internal part.
For our purpose it is convenient to develop a diagrammatic representation for the  elements
of the matrix ${\mathcal K}^{-1}$ and its powers. We denote the powers $\beta_0^n(\omega)$ of the diagonal elements of
${\mathcal K}^{-1}$ as empty circle, $\bigcirc$, elements $\beta_{1}(\omega,-w)$ and $\beta_{1}(\omega,w)$ as left and right arrows
from $\omega$ to $\omega-w$ and $\omega+w$ respectively. Similar arrows, but between $\omega$ and $\omega\pm 2 w$,
will represent elements $\beta_{2}(\omega,\pm w)$ (see Fig.1).

\begin{figure}[t]
\centerline{\includegraphics[width=75mm,angle=0,clip]{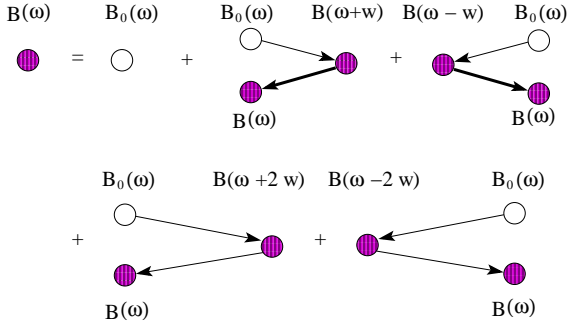}}
\caption{Dyson type equation for the diagonal elements of the matrix $ {\mathcal K}$ (full circles; see Fig.~3).
Each of the five diagrams contributes to diagonal matrix element of ${\mathcal K}$. Thin arrows represent bare
``hopping'' elements $\beta_1(\omega,\pm w)$ and $\beta_2(\omega,\pm w)$. Thick arrows represent
corresponding dressed  ``hopping'' elements (see Fig.~4). }
\label{fig:2}
\end{figure}

Matrix elements of ${\mathcal K}$ can be
found by analyzing the structure of powers of $1-{\mathcal K}^{-1}$.
They have the following form
\bea
\label{inverse-1}
{\mathcal K}(\omega,\omega^\prime)&=&B(\omega,w)\delta(\omega-\omega^\prime)\\
&+&\sum_{n=1;\sigma=\pm}^{\infty}\Gamma_n(\omega,\sigma w)\delta(\omega+\sigma n w-\omega'),\nn
\eea
where diagonal, $B$,
and off-diagonal, $\Gamma_n$, elements of ${\mathcal K}$ will be determined below [see Eq.~(\ref{Bn})].
Importantly, Fourier transform of this expression yields the
correlation function, $K(t,t')$. Diagonal and off-diagonal elements of ${\mathcal K}$ satisfy certain functional relations.
We identify these relations below and solve them exactly.

\begin{figure}[t]
\centerline{\includegraphics[width=75mm,angle=0,clip]{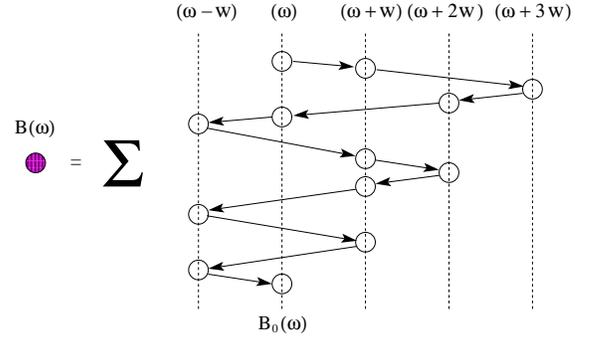}}
\caption{Diagrammatic representation for diagonal elements of the matrix $\mathcal{K}$.
}
\label{fig:11}
\end{figure}

\begin{figure}[t]
\centerline{\includegraphics[width=80mm,angle=0,clip]{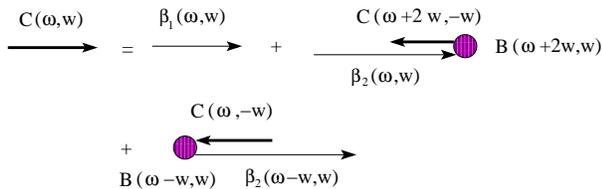}}

\caption{Dyson type equation for the non-diagonal (hopping) elements of the matrix ${\mathcal K}$.}
\label{fig:3}

\end{figure}

Let us start with
diagonal elements of ${\mathcal K}$, namely, with the elements of the form $B(\omega,w)\delta(\omega-\omega')$.
These terms define the translational invariant part of $K(t,t')$, which in time
representation depends on the difference, $t-t'$, due to the presence of $\delta(\omega-\omega')$ in (\ref{inverse-1}).
One can write Dyson type equation for the diagonal elements represented diagrammatically
in Fig.~\ref{fig:2}, where the full circles and the thick lines
represent full series of diagrams presented in Figs.~\ref{fig:11} and \ref{fig:3} respectively,
and the empty circle, marked as $B_0(\omega,w)$, represents the full sum of bare diagonal elements
of the form $\left[1+2 \omega-\beta_0(\omega,w)\right]^k$, leading to
\bea
\label{dyson-0}
\!\!B_0(\omega,w)=\sum_{k=0}^{\infty}\left[1+2 \omega-\beta_0(\omega,w)\right]^k=\frac{1}{\beta_0(\omega,w)-2 \omega}.\nn\\
\!\!\!\!\!\!\!\!\!\!\!
\eea
Functional relations pertinent to the diagrammatic series of Figs.~\ref{fig:2} and ~\ref{fig:3}
read
\bea
\label{dyson-1}
&&B(\omega,w)=B_0(\omega,w)-B_0(\omega,w)B(\omega,w)\\
&\times&\sum_{k=1,2;\sigma=\pm}
\beta_k(\omega,\sigma w)C(\omega+\sigma w,-\sigma w)B(\omega+\sigma k w,w).\nn
\eea
Similarly, the off-diagonal terms in Eq.~(\ref{inverse-1}), that contain $\delta(\omega=\omega'\pm w)$ satisfy the relations
following from Fig.~\ref{fig:3}:
\bea
\label{dyson-2}
&&C(\omega,w)=-\beta_{1}(\omega,w)-\beta_{2}(\omega,w)C(\omega+2 w,-w)\\
&&\times B(\omega+2 w,w)-C(\omega,-w) \beta_{2}(\omega-w,w)B(\omega-w,w).\nn
\eea
In Eqs.~(\ref{dyson-1}) and (\ref{dyson-2}), $C(\omega,\pm w)$ represents the fully dressed
hopping matrix element of $(1-{\mathcal K}^{-1})^{-1}$, that corresponds to the transition from
$\omega$ to $\omega \pm w$.

Finally, all remaining terms, $\Gamma_n(\omega,w)$, in Eq.~(\ref{inverse-1}) can be expressed via
$B(\omega + n w,w)$ as follows
\bea
\label{Bn}
\Gamma_n(\omega,w)=B(\omega,w)\prod_{k=1}^{n} C[\omega\pm (k-1) w]B(\omega\pm k w,w).\nn\\
\eea

%

Solving Eq.~(\ref{dyson-1}) with respect to $B(\omega,w)$, one obtains it in terms of  $B(\omega \pm w,w)$.
This relation provides a possibility to generate continued fraction form of the solution
for $B(\omega,w)$. Iterations in  Eqs.~(\ref{dyson-1}) and (\ref{dyson-2}) lead to the relations
\bea
\label{iteration-1}
\!\!\!\!&&\!\!\!\!C_{0}(\omega,w)=-\beta_{1}(\omega,w),\\
\!\!\!\!&&\!\!\!\!C_{m+1}(\omega,w)=-\beta_{1}(\omega,w)-C_{m}(\omega+w,-w)\beta_{2}(\omega,w)\times\nn\\
\!\!\!\!&&\!\!\!\!B_{m}(\omega+2 w,w)
-C_{m}(\omega,-w)\beta_{2}(\omega-w,w)B_{m}(\omega - w,w),\nn\\
\!\!\!\!&&\!\!\!\!B_{m+1}(\omega,w)^{-1}=-2 \omega+\beta_0(\omega,w)\nn\\
\!\!\!\!&&\!\!\!\!-\!\!\sum_{k=1,2; \sigma=\pm}\beta_k(\omega,\sigma w)C_{m}(\omega +\sigma w,-\sigma w)B_{m}(\omega+\sigma w,w).\nn
\eea
While the solution of Eqs.~(\ref{dyson-1}) and (\ref{dyson-2}) is simply defined by
the infinite number of iterations
$B(\omega,w)=\lim_{n\rightarrow\infty}B_{n}(\omega,w),\qquad
C(\omega,w)=\lim_{n\rightarrow\infty}C_{n}(\omega,w)$,
which is the continued fraction representation of the diagonal and off-diagonal matrix elements of
${\mathcal K}(\omega,\omega^\prime)$. Analytical expressions for diagonal elements of ${\mathcal K}$
obtained within two- and three-iteration approximation are presented in Appendix~B.

\begin{figure}[t]
\centerline{\includegraphics[width=85mm,angle=0,clip]{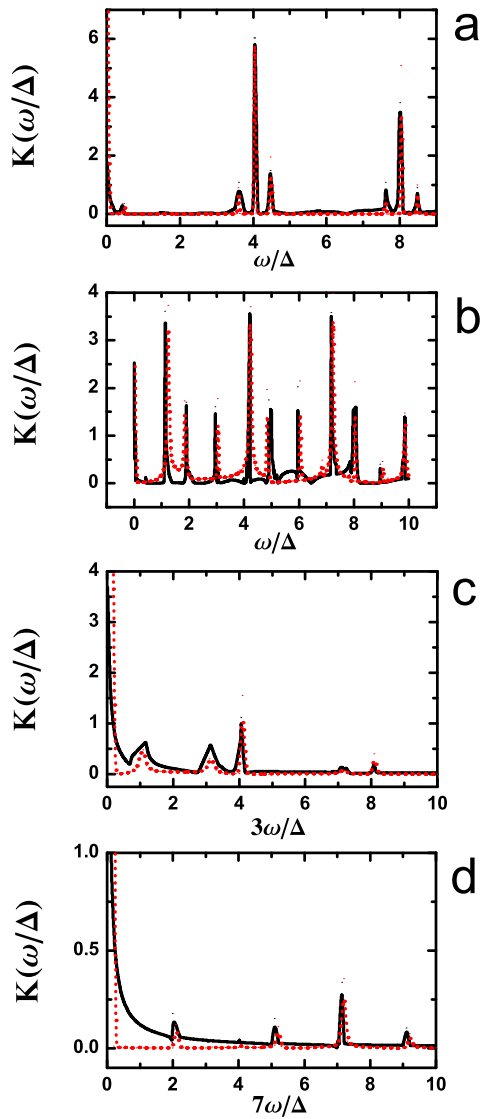}}
\caption{(Color online) Absolute value of the nonlinear dielectric response function plotted vs $\omega/\Delta$ for
various values of parameters. Full line is plotted from analytical expressions (see Eq.~(\ref{inverse-1}) and below),
while dotted line is obtained from exact numerics. Positions of the peaks determine the frequencies of the oscillatory $P_{\downarrow,\downarrow}(t,t^{\prime})$.
From top to bottom: a. $w/\Delta=4$, $\varepsilon/\Delta$=4, and $A/\Delta$=2.05;
b. $w/\Delta=3$, $\varepsilon/\Delta=4$, and $A/\Delta=2.05$; c. $w/\Delta=4/3$, $\varepsilon/\Delta$=1/3, and $A/\Delta=1/6$;
d. $w/\Delta=2/7$, $\varepsilon/\Delta$=1/7, and $A/\Delta$=1/28. Note that parameters corresponding to figures c and d
are out of the reach of the methods used in Ref.~\onlinecite{Hausinger2010}.}
\label{Cond}
\end{figure}

Evaluating the matrix elements of $K(\omega)=\int d\omega^\prime {\mathcal K}(\omega,\omega^\prime)$, which is
the Fourier image
of the correlation function $K(t,0)$, we find the spectrum of frequencies
which contribute here.  It is clear that all the matrix elements,
 $\Gamma_n(\omega,\pm w)$, of ${\mathcal K}(\omega,\omega^\prime)$, defined by Eq.~(\ref{Bn}),
will contribute to $K(\omega)$ substantially.
It is also clear that due to the periodicity of $B(\omega,w)+\sum_{n=1}^{\infty} \Gamma_n(\omega,\pm w)$, as a function of $\omega$,
which defines the Fourier image
of the correlation function, $K(t,t^\prime)$, only a part of terms in the sum
give essential contribution in the particular region of $\omega$. Contribution of the tail becomes
progressively smaller. We analytically calculate the first four elements of this sum, which gives the main contribution
into the spin-spin correlation function in the region $0 < \omega/\Delta < 10$
for the particular choices of the parameters $\Delta$, $w$, $a$, and $A$. We have restricted ourselves
within the fourth iteration level of the solution of Eqs.~(\ref{dyson-1}) for the same values of parameters.
This means that we cut the exact continued fraction representation of ${\mathcal K}(\omega,\omega^\prime)$ after four fractions,
as we checked that five and more iterations do not affect the result for these parameters in the plotted range
of $\omega/\Delta$.  Comparison of our analytical expression for $K(\omega)$  with
numerically evaluated solution of the corresponding Schr\"{o}dinger equation
are presented in Fig.~\ref{Cond} for various values of model parameters.
Note that in these plots we took into account
a finite relaxation rate, $\tilde\gamma$, in the Majorana fermion Green's function, which is needed to ensure the
causality. More specifically we chose $\tilde\gamma / \Delta\sim 10^{-2}$ to ensure exponential decay of the Green's
function at times $t>\tilde\gamma ^{-1}$.

\section{Summary}

In summary, we have developed a new technique for studies of general non-equilibrium two-level systems.
The technique is based on the mapping to a Majorana fermion field theory coupled to a scalar field.
We have applied the technique to study the dynamics of two-level systems with driving fields given by Eq.~(\ref{1}).
 Our analytical result for the nonlinear dielectric response
function in energy space is shown to be in good agreement with the numerically evaluated solution
of the time-dependent Schr\"{o}dinger equation.
We see that positions of the resonance peaks in $K(\omega)$
are in agreement with the results of exact numerical simulations.

Our technique allows generalization to {\em dissipative}  two level systems. We can consider decoherence
being caused by the dissipative environment and also generated by dissipative elements in superconducting
electronic circuits elements. One can extended our approach and include relaxation and dephasing times ($T_1$ and $T_2$)
into consideration. However, these problems as well as the comparison of our approach to the well known
comprehensive method based on the Bloch-Redfield equations~\cite{Slichter} is a subject of further
studies.

\begin{acknowledgments}

{\it Acknowledgments} -- This research was supported by the Intelligence Advanced
Research Projects Activity (IARPA) through the
US Army Research Office award W911NF-09-1-0351.

\end{acknowledgments}

\section{appendix A}


Here we will present some of the basic properties of Majorana fields, and will
calculate the Green's function of a free non-interacting Majorana fermions.
As fields, Majorana fermions are described by real Grassmann variables, $\xi$, with following
properties~\cite{Berezin,BM}
\bea
\label{grassmann}
\{\xi_{\mu}, \xi_{\nu}\}=0, \qquad \xi_{\mu}^+=\xi_{\mu},
\eea
where curly brackets stand for an anticommutator.

Integration rules over Grassmann variables are very simple. Namely
\bea
\label{grassmann-2}
\int d \xi_{\mu}=0,\;\;\; \int d \xi_{\mu} \xi_{\nu}=\delta_{\mu \nu}.
\eea
Using these rules of integration, one can prove that a Gaussian integral over
Majorana fermions is expressed in terms of the determinant of the quadratic form.
But contrary to ordinary fermionic integrals, it is equal to the square root of the
determinant:
\bea
\label{gauss}
\int {\cal D}\xi \exp\Big\{ \sum_{\mu,\nu=1}^N \xi_{\mu} A_{\mu\nu}\xi_{\nu}\Big\}=\sqrt{Det[A_{\mu\nu}]}
\eea
This expression is  correct both, for matrices and for differential operators.

In order to calculate the Green's function, $\langle \xi_{\mu}(t)\xi_{\nu}(t') \rangle$,
of non-interacting Majorana fermions, we introduce a generating functional
\bea
\label{gf}
{\cal Z}(\eta_{\mu})=\int{\cal D}\xi_{\mu} \exp\Big\{\int dt \Big[\frac{1}{4}\xi_{\mu}\dot{\xi}_{\mu}+\xi_{\mu}\eta_{\mu}\Big] \Big\}.
\eea
From this expression it is clear that
\bea
\label{gf1}
\langle \xi_{\mu}(t)\xi_{\nu}(t') \rangle = \frac{1}{{\cal Z}_0}\frac{\partial^2 {\cal Z}(\eta_{\mu})}{\partial\eta_{\nu}(t)\partial\eta_{\mu}(t')}{\Bigg|}_{\eta_{\mu}=0}.
\eea
So we need to calculate ${\cal Z}(\eta_{\mu})$. Calculation is straightforward.
One can find from Eq.~(\ref{gf})
\bea
\label{gf2}
{\cal Z}(\eta_{\mu})&=&\int{\cal D}\xi_{\mu}\\
&\!\!\times\!\!&\exp\Big\{\int dt \Big[\frac{1}{8}(\xi_{\mu}+ 2\eta_{\mu}d_t^{-1}) d_t ({\xi}_{\mu}- 2 d_t^{-1}\eta_{\mu})\nn\\
&\!\!+\!\!&\frac{1}{8}(\xi_{\mu}-2 \eta_{\mu}d_t^{-1}) d_t ({\xi}_{\mu}+2 d_t^{-1}\eta_{\mu})\Big]
- \eta_{\mu}d_t^{-1} \eta_{\mu} \Big\},\nn
\eea
where $d_t^{-1}=\frac{1}{2}sign[t-t']$ is the Green's function of the differential operator,
 $d_t\equiv\frac{d}{dt}$. According to (\ref{gauss}), the Gaussian integral over $\xi_{\mu}\pm 2\eta_{\mu}d_t^{-1}$
yields ${\cal Z}_0=\sqrt{Det[d_t]^2}$, which is a $C$-number, and we obtain
${\cal Z}(\eta_{\mu})= {\cal Z}_0 \exp\left\{- \eta_{\mu}d_t^{-1} \eta_{\mu} \right\}.$
Now, differentiating ${\cal Z}(\eta_{\mu}) $ twice with respect to $\eta_{\mu}$, and taking the limit $\eta_\mu=0$,
we reproduce the formula from the main text:
\bea
\label{gff}
\langle\xi_{\mu}(t_1) \xi_{\nu}(t_2)\rangle &=&\frac{1}{{\cal Z}_0}\int {\cal D}{\bm \xi}(t) \xi_{\mu}(t_1) \xi_{\nu}(t_2)\\
&\times&\exp \left[\frac{1}{4}\int dt \; \xi_{\mu} \dot{\xi}_{\mu}\right]
=\delta_{\mu\nu}\text{sign}[t_1-t_2].\nn
\eea

\section{Appendix B}

In this Appendix we present approximate expressions for solution of
Eqs. (\ref{dyson-1}) and (\ref{dyson-2}), which where obtained within
two and three iterations ({\em i.e.} by cutting the continued fraction representation after
two and three fractions).

Within two-iteration approximation  we have for the  diagonal element, $B_1(\omega,w)$, of matrix
${\cal K}$:
\begin{widetext}
\bea
\label{iter-2}
B_1(\omega,w)=\qquad\qquad\qquad\qquad\qquad\qquad\qquad\qquad\qquad\qquad\qquad\qquad\qquad\qquad\qquad\qquad\qquad\qquad\qquad\qquad\qquad\\
\frac{1}{-2 \omega+\beta_0(\omega,w)+\frac{\beta_1(\omega,w)}{-2 (\omega+w)+\beta_0(\omega,\omega+w)}
+\frac{\beta_1(\omega,-w)}{-2 (\omega-w)+\beta_0(\omega,\omega-w)}+\frac{\beta_2(\omega,w)}{-2 (\omega+2 w)+\beta_0(\omega,\omega+2 w)}
+\frac{\beta_2(\omega,-w)}{-2 (\omega-2 w)+\beta_0(\omega,\omega-2 w)}}.\nn
\eea
Similarly, for hopping element $C_1(\omega,w)$ we get from Eq.~(\ref{iteration-1})
\bea
\label{iter-22}
C_1(\omega,w)=-\beta_{1}(\omega,w)+\frac{\beta_{2}(\omega,w)\beta_1(\omega+2 w,-w)}{-2(\omega+2w)+\beta_0(\omega+2w,w)}
+\frac{\beta_1(\omega,-w) \beta_{2}(\omega-w,w)}{-2(\omega-w)+\beta_0(\omega-w)}.
\eea
This yields the following expression for diagonal elements of ${\cal K}$ in a three-iteration approximation
\bea
\label{iter-3}
B_2(\omega,w)=\qquad\qquad\qquad\qquad\qquad\qquad\qquad\qquad\qquad\qquad\qquad\qquad\qquad\qquad\qquad\qquad
\qquad\qquad\qquad\qquad\qquad\qquad\qquad\nn\\
\frac{1}{-2 \omega+\beta_0(\omega,w)-C_1(\omega,w)B_1(\omega,\omega+w)
-C_1(\omega,-w)B_1(\omega,\omega-w)+\beta_2(\omega,w)B_1(\omega,\omega+2 w)
+\beta_2(\omega,-w)B_1(\omega,\omega-2 w)}.\nn
\eea

Finally, within the same level of approximation one has for hopping elements $C_2(\omega,w)$:
\bea
\label{iter-33}
C_2(\omega,w)=-\beta_{1}(\omega,w)-\beta_{2}(\omega,w)C_1(\omega+2 w,-w)B_0(\omega+2 w,w)
-C_1(\omega,-w) \beta_{2}(\omega-w,w)B_0(\omega-w,w).
\eea
As we see from Eq.~(\ref{iteration-1}), one can continue this procedure to higher levels of iteration to analytically calculate
the survival probability of the spin with an arbitrary accuracy.
\end{widetext}




\end{document}